# Using a Nature-based Virtual Reality Environment for Improving Mood States and Cognitive Engagement in Older Adults: A Mixed-method Feasibility Study

**Saleh Kalantari[1*] , Tong Bill Xu[1], Armin Mostafavi[1], Angella Lee[1], Ruth Barankevich[1], Walter Boot[2], Sara Czaja[3]**

[1] Department of Human Centered Design, Cornell University, Ithaca, NY, USA
[2] Department of Psychology, Florida State University, Tallahassee, FL, USA
[3]Center on Aging and Behavioral Research, Division of Geriatrics and Palliative Medicine, Weill Cornell Medicine, New York, NY, USA

Corresponding author: Saleh Kalantari <sk3268@cornell.edu>

## Abstract

Engaging with natural environments and representations of nature has been shown to improve mood states and reduce cognitive decline in older adults. The current study evaluated the use of virtual reality (VR) for presenting immersive 360 degree nature videos and a digitally designed interactive garden for this purpose. Fifty participants (age 60 plus), with varied cognitive and physical abilities, were recruited. Data were collected through pre/post-intervention surveys, standardized observations during the interventions, and post-intervention semi structured interviews. The results indicated significant improvements in attitudes toward VR and in some aspects of mood and engagement. The responses to the environment did not significantly differ among participants with different cognitive abilities; however, those with physical disabilities expressed stronger positive reactions on some metrics compared to participants without disabilities. Almost no negative impacts (cybersickness, task frustration) were found. In the interviews some participants expressed resistance to the technology, in particular the digital garden, indicating that it felt cartoonish or unappealing and that it could not substitute for real nature. However, the majority felt that the VR experiences could be a beneficial activity in situations when real-world contact with nature was not immediately feasible.

**Using a Nature-based Virtual Reality Environment for Improving Mood States and Cognitive Engagement in Older Adults: A Mixed-method Feasibility Study**

## 1. Introduction

Due to extended lifespans and an aging population, one in five adults in the United States will reach retirement age by the year 2030 (Schwenk et al., 2016; Vespa et al., 2020). The growth in this segment of the population has prompted healthcare and technology sectors to focus more strongly on the well-being and specific health needs of older adults. These needs are in many cases related to declines in physical performance (Makizako et al., 2017) and cognitive abilities (Murman, 2015) that can sometimes result in a loss of independence (Lau et al., 2015). Millions of Americans older adults currently suffer from cognitive impairment (CI), a condition that can result from a variety of underlying causes, including both memory-related and non-memory related losses of function, and that often progresses to more severe forms of dementia (Roberts & Knopman, 2013; Lee et al., 2014). In addition to the direct impacts for individuals who are living with CI, family members and caretakers often confront burdens associated with the effects of the condition (Lee et al., 2014; Park & Shin, 2016). Cognitive impairment may also contribute to the high numbers of older adults who struggle with loneliness (Ong et al., 2016; Perissinotto et al., 2012; Yang & Victor, 2011) and depression (Broadhead et al., 1990; Fiske et al., 2009; Kennedy et al., 1991; Unützer et al., 1997). In recent years, the effects of the COVID-19 pandemic have added to the challenges of older adults living with CI, as the need for restrictions on social contact has reduced opportunities for engagement and travel, as well as limiting the accessibility of in-person healthcare interventions (Macdonald & Hülür, 2021; Paananen et al., 2021). Fortunately, there are a wide range of other options available to help mitigate the effects of CI. These interventions are rapidly shifting away from an exclusive focus on in-person care to incorporate technological aids, telemedicine, video-based socializing, and other care and activity programming grounded in electronic media (Steinman et al., 2020).

### *1.1. Exposure to Nature as an Intervention for People with Cognitive Impairment*

Engaging with natural environments has been linked to improvements in cognitive functioning. For example, researchers have found that increases in neighborhood vegetation are associated with slower cognitive decline (de Keijzer et al., 2018), a lower prevalence of Alzheimer's Disease (Brown et al., 2018), and overall better cognitive abilities among older adults (Prohaska et al., 2009). Improvements in physiological factors such as immune function, blood pressure, and heart rate have also been linked to nature exposure, which may help to explain its impact on cognitive functioning (Alfonsi et al., 2014). Many researchers have also focused on the mental health benefits of nature, particularly in regard to reductions in anxiety, depression, apathy, and negative mood states, as a hypothesized mechanism for its positive impact on cognitive function (Barnes et al., 2006; Besser, 2021; Chalfont et al., 2020; Heath, 2004; Kaplan, 1995; Ma, 2020; Pun et al., 2018). The benefits of nature exposure for improving mood states has been widely documented in diverse contexts, ranging from wilderness areas to residential streets to urban green spaces and gardens (Aspinall et al., 2015; Brooks et al., 2017; Gidlow et al., 2016; Kondo et al., 2018; Mayer et al., 2009; Song et al., 2013, 2015; Ulrich, 2002; Van Den Berg & Custers, 2011). Since benefits for mood states and cognitive function have been found even in "micro-doses" of nature such as viewing urban gardens, the incorporation of green spaces into the built environment could potentially have widespread impacts in the prevalence and severity of cognitive impairment. However, the rapid increase in urbanization globally has left many city-dwellers without access to such spaces (Kalantari & Shepley, 2020; Patricio et al., 2020; *World Urbanization Prospects*, 2019). Furthermore, even when they do have proximal access to nature, older adults are often limited in their ability to experience such spaces by obstacles such as limited mobility, pain, and fear of falling (Appel et al., 2020).

### *1.2. Virtual Reality and Nature*

When direct access to nature is not available on a regular basis, researchers and healthcare practitioners have turned to supplementary approaches for obtaining some of its benefits, such as window views (Chang & Chen, 2005; R. Kaplan, 2001; Raanaas et al., 2016), nature-oriented artwork and murals (Diette et al., 2003), nature videos (Kahn et al., 2008), and virtual reality immersion (Gorini et al., 2010; Moyle et al., 2018). Like all such approaches, the use of virtual reality (VR) has liabilities, particularly in regard to its lack of tactile engagement and its inability to fully replicate the deep complexity and material interconnectedness of actual organic environments (Kalantari & Neo, 2020). At the same time, however, the use of VR headsets—or, less commonly, full-room projection systems (Annerstedt et al., 2013; Wang et al., 2020)—helps to create an immersive experience that reduces distractions and can grant reprieve from negative features of the immediate real-world context (Furman et al., 2009; Gorini et al., 2010). VR is also a very cost-effective intervention in comparison to the construction of real-world green spaces and/or supervised travel to those spaces (Yeo et al., 2020), and it can promote active engagement through the incorporation of game-like features (Kalantari, 2019). While the use of VR should not be seen as a substitute that replaces the imperative toward real-world biophilic and sustainable design, it can nonetheless provide an effective supplementary approach, especially in situations where direct access to nature is not immediately feasible (Kalantari et al., 2021).

The most basic form of nature exposure through VR focuses on experiencing immersive video footage, such as vistas, beaches, forests, and gardens (Anderson et al., 2017; Appel et al., 2020; Gorini et al., 2010; Moyle et al., 2018; Nukarinen et al., 2020; Riva et al., 2020; Yeo et al., 2020). Such experiences may involve a static user position or a pre-determined movement path, while allowing users to freely look around and view different portions of the surroundings through the use of head-motion-tracking technologies. This type of VR is relatively simple and low-cost to construct, as all it really requires is multi-directional video-recording equipment (Nukarinen et al., 2020). Studies using these methods have found improvements in relaxation, mood, and alertness during and after the VR experiences of nature (Anderson et al., 2017; Moyle et al., 2018). However, this type of experience is still quite passive, as the videos generally lack opportunities for engagement and exploration. Appel and colleagues (2020) noted that once participants had experienced a VR video-footage scene in its entirety, they showed little interest in immediately viewing it again or looking for additional details. As such, the level of cognitive engagement was not much different from watching a flat-screen video. While the passive viewing of VR nature scenes does have demonstrated benefits for mood improvement and stress-reduction, the technology is available to create more interactive components and thus prompt greater active engagement. Rendering programs can be used to create digital scenes and elements that are programmed to respond to user's actions, and that can allow users to fluidly move through the environment following a path of their own choosing. This approach has not yet been widely applied or studied in the context of mental health or cognitive function.

### *1.3. Virtual Reality and Older Adults*

VR technologies have previously been tested with older populations and found to be an effective tool for such individuals. A review by D'Cunha and colleagues (2019) found strong evidence that older adults who used VR tended to experience mood improvements and reductions in apathy, and that the participants reported enjoying the experiences. Manera and colleagues (2016) found that participants in a reminiscence-focused intervention using VR reported greater satisfaction with the experience compared to those who completed a similar paper-based activity. Appel and colleagues (2020) gathered feedback from older adults, some of whom had been diagnosed with mild cognitive impairment, on the experience of viewing brief nature scenes (less than 10 minutes) presented using VR. The responses to these experiences were quite positive, with a high reported comfort level and enjoyment, and no reported negative side-effects. Researchers have also found that attitudes toward VR among older adults significantly improved after experiencing the technology for the first time (Huygelier, Schraepen, van Ee, et al., 2019). Nonetheless, negative impacts of VR have occasionally been reported in the wider literature, including experiences of cybersickness, headaches, and very rarely, the triggering of epileptic seizures (Keshavarz et al., 2018; Seifert & Schlomann, 2021).

### *1.4. Study Goals and Hypotheses*

The present study used an active-engagement approach to nature-based VR interventions for older participants, including participants with different cognitive capabilities (ranging from a cognitive impairment to no cognitive impairment). The researchers designed, implemented, and tested a new digital "VR Garden" in which users can interact with plants and animals and engage in gardening activities. We also include a module with passive 360-degree nature videos. In the current study the emphasis was on the feasibility, usability, and likely adoption of these VR environments. We employed a variety of quantitative measurement instruments (discussed in more detail below) to evaluate mood, engagement, and perceptions of the technology. We also gathered qualitative/interview data to obtain further insights about the participants' experiences. The current study did not address the long-term impacts of the VR garden on cognitive function; this will be evaluated in future work. The study was designed to test four primary hypotheses:

*Hypothesis 1*: Short-term exposure to the Virtual Garden will be associated with improvements in mood states.

*Hypothesis 2*: Short-term exposure to the Virtual Garden will be associated with improvements in attitudes toward VR technologies.

*Hypothesis 3*: Changes in mood states and changes in attitudes toward VR technology will be similar between participants who have mild cognitive impairment vs. those who do not have cognitive impairment.

*Hypothesis 4*: Positive changes in mood states and in attitudes toward VR technology will be significantly more pronounced among participants who have physical disabilities, compared to those who do not have physical disabilities.

## 2. Methods

### *2.1. Design of the Virtual Environment*

The researchers designed the Virtual Garden using modeling, UV-mapping, and mesh optimization in Autodesk 3ds Max. The design documents were imported into Epic Games' Unreal Engine 4, and blueprint visual scripting along with C++ scripting was then used to create the interaction components. We included four modules within the overall VR environment: a tutorial, a set of 360-degree nature videos, an interactive garden, and a gardening game.

The *tutorial* module was designed as a learning experience to help participants become familiar with the VR system and its navigational and interaction controls. In this module the participants encountered a relatively small (40x40 meters) open space surrounded by walls. The space included several distinct interaction areas, highlighted in different colors, where participants could read instructional text and learn and practice with the VR controls. Upon entering the area, the participants were asked to learn how to move themselves to each interaction site, and then to follow the instructions and interact with the virtual objects located at each site. All of the skills needed to navigate the Virtual Garden were practiced in the learning trial, including moving from place to place at different speeds, interacting with textual elements, and grabbing and moving objects (Figure 1a).

The *nature videos* module presented a passive restorative experience. We used a GoPro Fusion 360 camera to record nine short videos of natural areas and botanical gardens in the Ithaca, NY area. The resolution of these videos was set to 5k and then later downsized to match the VR display capability, as discussed below. The videos were between 30–45 seconds in length, for a total combined duration of about 5 minutes. They were rendered using Adobe Premiere Pro CC and the GoPro FX Reframe plugin, and then projected onto a sphere as an environment within the Unreal Engine. Participants were limited to a static position when viewing these videos, but they were able to freely look around and view different perspectives on the 360-degree environment. The videos included recorded sounds from the respective natural areas (Figure 1b).

The *interactive virtual garden* was an artificial environment created by the researchers to reflect experiences of nature while promoting active engagement. The design of the engagement components drew heavily on existing evidence-based design guidelines for therapeutic gardens (Marcus & Sachs, 2013). Participants were able to experience the virtual garden in a fairly passive fashion by simply "walking" through it (using hand-held controllers to initiate or pause motion along designated paths) and looking around to observe the trees, plants, flowers, ponds, fountains, and benches along the way. They could also engage in interactions with various elements of the garden, for example by touching the flowers (in response butterflies would come out of some areas), feeding the ducks in the pond, and throwing rocks into the pond. The vibration functions of the hand-held controllers as well as natural sound recordings were used as feedback for interactions and to help improve the garden's realism (Figure 1c).

The *gardening game* was the most interactive aspect of the virtual experience. In this module participants were able to create a garden layout and plant and water a variety of flowers to achieve their own desired aesthetic. Eight cultivation areas were included in the overall garden, each containing multiple planting spots. Participants were able to choose from different flowering plants (Gazania, Antirrhinum, Salvia, Narcissus, Petunia, Papaver, Lilium, and Eschscholzia) that they could move from a storage area to one of the planting spots. To minimize any physical difficulties related to bending down, the planting process was designed as an automatic drop from a hand icon into the soil, complete with a rewarding sound when the planting was accomplished. The participants could then use a watering can to nurture their garden, which resulted in the growth of the plants (Figure 1d). The goal of this mini-game within the virtual garden was to promote engagement and the maintenance of cognitive skills (attention, memory, and executive control) (Anguera et al., 2013).

### *2.2. VR Presentation Equipment*

To present the VR environments, we used a consumer version of the Oculus Quest 2 headset with Oculus Touch controllers for the right and left hands, for all sessions and all participants. The resolution of the Oculus Quest 2 is 1832×1920 per eye, with a 90 Hz refresh rate. The head-mounted display has a six-degrees-of-freedom inside-out tracking system, which uses external references in the real-world environment to precisely determine the direction of the user's gaze. The participants could choose to be seated or standing during all interactions, and motion within the VR environment was enacted through the hand controllers. The camera height within the environment was determined automatically based on each participant's eye-height above the floor (Figure 2).

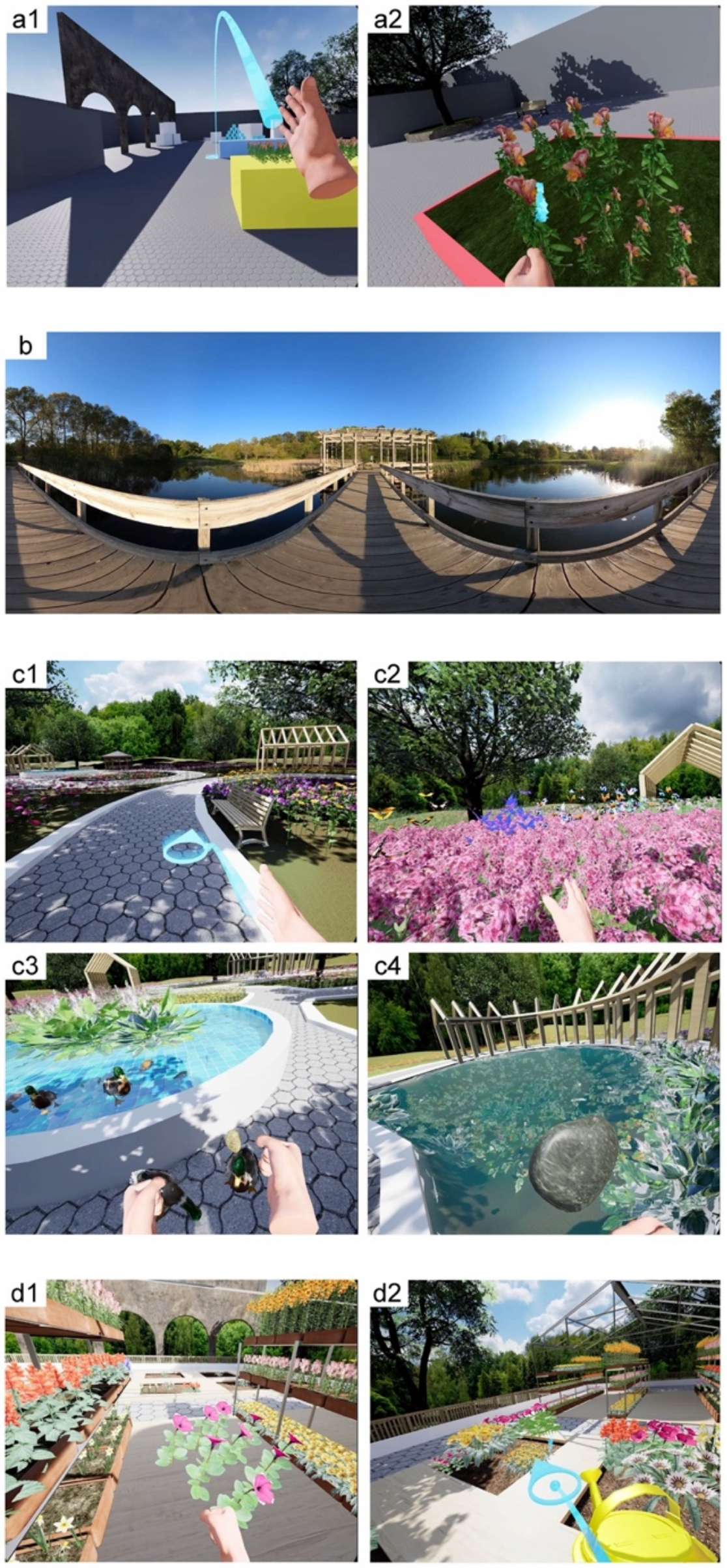

**Figure 1.** Screenshot captures of the VR Garden environments. (a) The tutorial helped participants to learn the VR controls. (b) The video module allowed participants to view short, 360-degree footage of local natural areas. (c) The interactive virtual garden allowed participants to engage in activities such as (c1) walking along pathways, (c2) touching the flowers, (c3) feeding the ducks, and (c4) throwing rocks into a pond. (d) The gardening game allowed participants to layout their own cultivation areas in which they could plant and water various types of flowers.

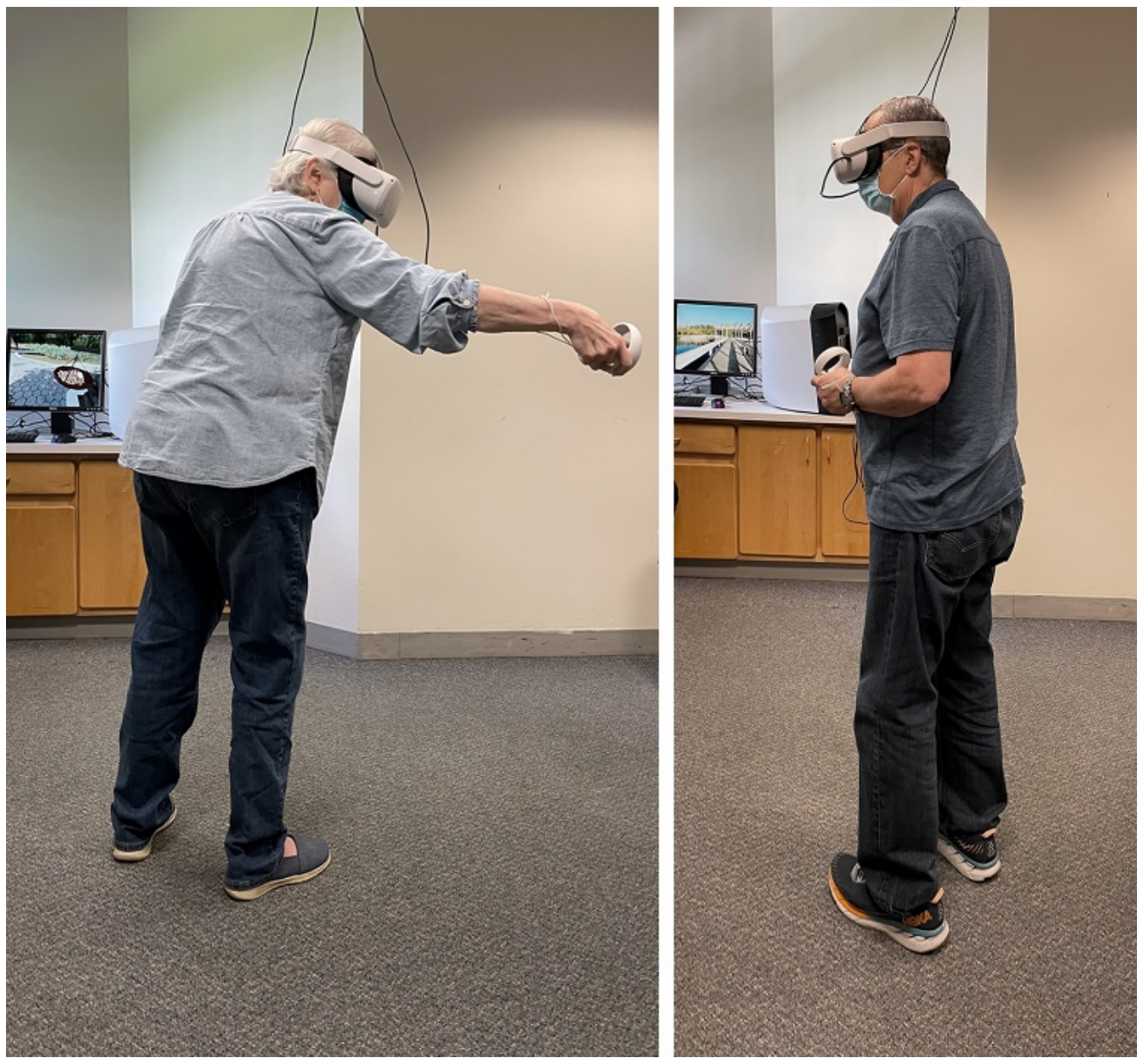

**Figure 2.** Examples of study participants interacting with the Virtual Garden (images used with the participants' permission).

*2.3.Participants and Recruitment*

An a-priori analysis of the required sample size for this study was conducted using the G*Power software tool. For an effect size of dz = 0.50, an alpha level of 0.01, and a statistical power of 0.80, the necessary sample size was 51 participants. The effect size of dz = 0.5 was estimated from Appel and colleagues' (2020) effect size for attitude changes toward VR technologies among older adults who experienced nature-oriented VR for the first time. We used a convenience sampling approach to recruit 52 participants in the Ithaca, NY area, using fliers in local senior living centers and calls for volunteers on community e-mail lists. All of the participants provided informed consent to participate in the study. The recruitment procedures and overall study protocols were reviewed and approved by the Institutional Review Board at Cornell University. We screened the participants for exclusion criteria focused on a history of seizures, epilepsy, severe motion sickness, and the use of implanted medical devices such as pacemakers. One participant was excluded from the study upon discovering that she used a pacemaker, and a second participant withdrew after experiencing a heightened confusion episode during the experiment. Data is reported for the remaining 50 participants.

*2.4.Study Procedure*

After providing informed consent each participant was asked to fill out a demographic questionnaire, which gathered information on age, gender, marital status, race/ethnicity, employment status, highest completed level of education, income level, experiences of depression, prior diagnosis of physical disability, proficiency with computers, and extent of exposure to nature in everyday life (the full questionnaire is included in Appendix One). This survey form was completed remotely and on the participant's own time. After filling out this initial form, each participant was invited to schedule an experimental session at the study site in Lifelong Institute in Ithaca, NY. These sessions were conducted individually. Upon arriving for the session, each participant filled out additional questionnaires to assess current mood states, attitudes toward virtual reality, and cognitive capabilities (Appendix One). The researchers then introduced the participant to the hand-held controllers and assisted with donning the VR headset. Two researchers were present at each session, one of whom focused on providing technological assistance and support for the participant, and the other of whom focused on conducting questionnaires and making empirical observations of engagement and expressed reactions throughout the experiment.

After donning the headset each participant first entered the tutorial module of the VR environment, which provided an opportunity to become familiar with the equipment and controls. This tutorial required approximately 5 minutes to complete, with very little variation in completion times or observed frustration levels among the different participants. After completing the tutorial each participant was asked to remove the headset and take a two-minute break, during which they engaged in small-talk and informal feedback with the researchers. Next, the participant was ask to experience the nature videos module (approximately 5 minutes), the interactive garden (approximately 10 minutes), and the gardening game (approximately 8 minutes). Between each of these sections the participant again removed the headset and took a two-minute break.

After completing all of the VR modules, each participant was asked to fill out a third questionnaire, which repeated the previous assessments of current mood and attitudes toward virtual reality, as well as assessments of immersion and cybersickness (Appendix Two). The researcher then conducted a semi-structured interview with each participant to gather qualitative feedback about their experience of the VR environment. With the participants' permission, these interviews were recorded and then transcribed for thematic analysis. Upon completing the session, each participant was presented with a small gift certificate and a visual snapshot of the virtual garden that they had created during the game module.

### *2.5.Measurement Tools*

The study used a wide array of measurement instruments derived from prior research. In the demographic survey, the *Nature Exposure Scale II* was used to assess each participant's exposure to nature in everyday life. This simple instrument consists of 6 items, each rated on a 5-point Likert scale, with 1="low" and 5="high" (Wood et al., 2019). The ratings on the 6 items were summed to obtain a total nature-exposure score with a possible range of 6–30.

Also on the demographic survey, the participants' likelihood of having depression was assessed using the *Center for Epidemiologic Studies Depression Scale* (Hann et al., 1999). This instrument consists of 10 items, each on a 4-point Likert scale, with positive items reverse coded (in other words, higher scores indicate a greater likelihood of having depression). Ratings on all of the items were summed, and participants with a total score greater than 16 were classified as likely to be suffering from some degree of depression.

Proficiency with computers was assessed in the initial demographic questionnaire using two scales: the *Computer Self-efficacy Scale* (Barbeite & Weiss, 2004) and the *Computer Proficiency Scale* (Boot et al., 2015). The first instrument focuses on confidence levels in relation to information technology, while the second instrument focuses on proficiency with different computer domains (printers, the Internet, entertainment software, etc.). Both instruments use 5-point Likert scales, with higher scores indicating greater proficiency.

Participants' cognitive capabilities were assessed at the time of the experiment using the *Montreal Cognitive Assessment* (Nasreddine et al., 2005). This instrument does not in itself confirm a diagnosis of cognitive impairment, but it is commonly used as a screening tool to assess global cognitive status. The assessment involves several brief written and verbal tasks measuring executive functions, memory, language, and reasoning. In the current study the assessment was used to divide the participants into those with a likely cognitive impairment and those without a cognitive impairment, with a threshold score of lower than 26 on the instrument designated as likely-impaired. There has been some debate regarding the exact threshold score on this instruments that indicates likely cognitive impairment, but > 26 is the most commonly used metric (Carson et al., 2018; Milani et al., 2018; Wong et al., 2015). None of the participants in the current study scored low enough on this instrument to be regarded as having severe cognitive impairment.

The mood states of each participant were assessed immediately before and immediately after the VR session, using the *Multidimensional Mood State Questionnaire* (Steyer et al., 1997). This instrument includes 30 items, each on a 6-point Likert scale, with higher scores indicating more positive mood states. The results are divided into three dimensions, including "good/bad" mood (GB), "calm/nervous" mood (CN), and "awake/tired" mood (AT). The total scores of a participant for each of these three dimensions were calculated separately.

Participants' attitudes toward VR technology were also assessed immediately before and immediately after the VR session. The instrument used for this purpose was a scale developed by Huygelier and colleagues (2019) for the specific purpose of evaluating *Acceptance of Head-mounted Virtual Reality in Older Adults*. It includes 18 items, each on a 5-point Likert scale, with higher scores indicating a more positive attitude toward the technology. The scores on all items were summed to obtain a total VR-attitude score.

During the VR session, positive and negative emotional responses to the virtual environment were measured observationally using the *Observed Emotion Rating Scale* (Lawton, Van Haitsma, & Klapper, 1999). This instrument is used to evaluate five emotional states: 1, 2, 3, 4, and 5. The observing researcher tracked the total amount of time in which each of these states was demonstrated by the participant.

Immersion levels were measured after the VR session using the *MEC Spatial Questionnaire* (Vorderer et al., 2004). This instrument is a validated, multidimensional measure of spatial presence and its components which include seven dimensions, each with a 4-item scale including Attention Allocation, Spatial Situation Model, Spatial Presence, Cognitive Involvement, Suspension of Disbelief, Domain Specific Interest, and Visual Spatial Imagery. Response options range from 1 (strongly disagree) – 5 (strongly agree).

Finally, two instruments were used after the VR session to measure potential negative impacts of the technology. Participants were asked to complete the *NASA Task Load Index* (Hart & Staveland, 1988) to evaluate perceptions of effort and frustration as related to engagement with the Virtual Garden. This instrument includes six subscales, including mental demand, physical demand, temporal demand, effort, performance, and frustration level each on a 5-point Likert scale, with higher scores indicating greater perceived task loads. Total scores for each participant were obtained by summing the responses for all items. The *Simulator Sickness Questionnaire* (Kennedy, Lane, Berbaum, & Lilienthal, 2009) was used to assess experiences of "cybersickness." This instrument includes 16 items on a 4-point Likert scale range from none to sever, with higher scores indicating greater levels of discomfort. The responses on all items were summed to get a total cybersickness score for each participant.

## 3. Results

### *3.1.Descriptive Statistics and Potential Confounding Variables*

The average age of the participants was 67.98. The majority were female (74%), White (90%), and had completed at least a Bachelor's-level education (80%). A broad range of income levels were represented. The average score on the *Center for Epidemiologic Studies Depression Scale* was 12.36 (SD=6.88), and 13 out of the 50 participants exceeded the threshold for being classified as likely to be experiencing depression. The average score on the *Computer Self-efficacy Scale* was 56.86 (SD=8.16), and on the *Computer Proficiency Scale* it was 25.91 (SD=3.39). The participants' level of exposure to nature in everyday life was quite high, with an average score on the *Nature Exposure Scale II* of 26.6 (SD=4.04) out of a total possible 30 points. The majority of the participants (80%) had not been previously diagnosed with a physical disability. The average score on the *Montreal Cognitive Assessment* was 25.88 (SD=2.81), with 24 participants (48%) exceeding the threshold for likely having a cognitive impairment. Overall, the demographic variables were well-distributed among those with a cognitive impairment (CI) and those without (Non-CI) and among disability vs. non-disability participants (Table 1).

Descriptive statistics for the remaining measurements are presented in Table 2. There was a very low amount of cybersickness reported, with an average score of 2.70 (SD=3.51). The results for the *NASA Task Load Index* indicated an average score of 2.12 (SD=0.52). Scores on the Mood States instrument and Attitudes toward VR Technology are discussed in the following sections in relation to the specific study hypotheses.

We conducted correlation test with Holm's correction to search for potential relations among some of the demographic variables. The only significant correlation was found between the *Computer Self-efficacy Scale* and the *Computer Proficiency Scale*, which is expected since both are measurements of familiarity with information technology (Table 3). We also found that Nature Exposure scores were significantly correlated with changes in attitudes to technology during the experiment ($t= 2.847$, $p= 0.007$), and to a lesser extent, with spatial self ($t=1.906$, $p= 0.064$). Education level was significantly correlated with changes in spatial action during the experiment ($t=-2.230$, $p= 0.031$), and to a lesser extent, with spatial self ($t=-1.889$, $p= 0.066$).

**Table 1.** Demographic characteristics of the study participants.

| Characteristic | Total (n = 50) | Non-CI (n = 26) | CI (n = 24) | Non-Disability (n = 40) | Disability (n = 10) |
|---|---|---|---|---|---|
| **Age** | 67.98 (4.85) | 67.46 ( 4.88) | 68.53 ( 4.87) | 67.60 (4.89) | 69.18 (4.77) |
| **Gender** | | | | | |
| Male | 13 (26.0) | 4 (15.4) | 9 (37.5) | 12 (30.0) | 1 (10.0) |
| Female | 37 (74.0) | 22 (84.6) | 15 (62.5) | 28 (70.0) | 9 (90.0) |
| **Ethnicity** | | | | | |
| White | 45 (90.0) | 23 (88.5) | 22 (92.7) | 36 (90.0) | 9 (90.0) |
| All other ethnicities (including mixed ethnicity) | 5 (10.0) | 3 (11.5) | 2 (8.33) | 4 (10.0) | 1 (10.0) |
| **Marital Status** | | | | | |
| Single | 7 (14.0) | 4 (15.4) | 3 (12.5) | 5 (12.5) | 2 (20.0) |
| Married | 25 (50.0) | 13 (50.0) | 12 (50.0) | 22 (55.0) | 3 (30.0) |
| Divorced | 11 (22.0) | 6 (23.1) | 5 (20.8) | 9 (22.5) | 2 (20.0) |
| Separated | 3 (6.00) | 1 (3.85) | 2 (8.33) | 1 (2.50) | 2 (20.0) |
| Widowed | 4 (8.00) | 2 (7.69) | 2 (8.33) | 3 (7.50) | 1 (10.0) |
| **Education Level** | | | | | |
| Doctorate degree | 9 (18.0) | 4 (15.4) | 5 (20.8) | 6 (15.0) | 3 (30.0) |
| Professional degree | 3 (6.00) | 3 (11.5) | 2 (8.33) | 3 (7.50) | 0 (0.00) |
| Master's degree | 18 (36.0) | 18 (69.2) | 7 (29.2) | 17 (42.5) | 2 (20.0) |
| Bachelor's degree | 10 (20.0) | 10 (38.5) | 4 (16.7) | 5 (12.5) | 4 (40.0) |
| Associate degree | 3 (6.00) | 3 (11.5) | 2 (8.33) | 3 (7.50) | 0 (0.00) |
| Some college credit/no degree | 7 (14.0) | 7 (26.9) | 4 (16.7) | 6 (15.0) | 1 (10.0) |
| **Total Annual Income** | | | | | |
| Less than $50,000 | 12 (24.0) | 6 (23.1) | 6 (12.5) | 7 (17.5) | 3 (30.0) |
| $50,000 - $69,999 | 12 (24.0) | 7 (26.9) | 5 (20.8) | 10 (25.0) | 0 (0.00) |
| $70,000 - $89,999 | 10 (18.0) | 5 (19.2) | 4 (16.7) | 8 (20.0) | 2 (20.0) |
| $90,000 - $119,999 | 6 (12.0) | 4 (15.4) | 2 (8.33) | 6 (15.0) | 4 (40.0) |
| Greater than $120,000 | 3 (6.00) | 1 (3.85) | 2 (8.33) | 3 (7.50) | 0 (0.00) |
| Prefer not to answer | 8 (16.0) | 3 (11.5) | 5 (20.8) | 6 (15.0) | 1 (10.0) |
| **Montreal Cognitive Assessment** | 25.88 ( 2.81) | 27.96 ( 1.43) | 23.62 ( 2.10) | 26.15 (2.26) | 24.50 (4.17) |
| **Disability** | | | | | |
| Yes | 10 (20.0) | 4 (15.4) | 6 (25.0) | — | — |
| No | 40 (80.0) | 22 (84.6) | 18 (75.0) | — | — |

**Table 2.** Descriptive statistics for data collected during the VR sessions.

| Measure | Total (n=50) | Non-CI (n=26) | CI (n=24) | Non-Disability (n=40) | Disability (n=10) |
|---|---|---|---|---|---|
| Simulator Sickness | 2.70 (3.51) | 2.73 (3.41) | 2.67 (3.68) | 2.67 (3.81) | 2.80 (1.99) |
| NASA-TLX | 2.12 (0.52) | 2.05 (0.50) | 2.19 (0.55) | 2.11 (0.53) | 2.15 (0.51) |
| Social Desirability | 9.08 (2.11) | 9.15 (2.05) | 9.00 (2.21) | 9.20 (2.10) | 8.60 (2.17) |
| Spatial Self-Location | 34.52 (5.66) | 35.23 (4.36) | 33.75 (6.82) | 33.70 (5.98) | 37.80 (2.20) |
| Spatial Possible Actions | 30.52 (4.67) | 30.96 (4.28) | 30.04 (5.11) | 30.07 (4.94) | 32.30 (2.98) |
| System Usability Scale | 72.06 (5.04) | 70.73 (4.92) | 73.50 (4.85) | 71.72 (5.33) | 73.40 (3.53) |
| **Mood: Awake/Tired (AT)** | | | | | |
| *Pre-session* | 46.70 (6.62) | 45.62 (7.16) | 47.88 (5.91) | 46.65 (6.67) | 46.90 (6.76) |
| *Post-session* | 47.24 (7.94) | 47.54 (8.45) | 46.92 (7.51) | 46.73 (8.04) | 49.30 (7.59) |
| *Delta* | 0.54 (6.89) | 1.92 (6.22) | -0.96 (7.39) | 0.07 (6.77) | 2.40 (7.40) |
| **Mood: Calm/Nervous (CN)** | | | | | |
| *Pre-session* | 47.44 (5.68) | 46.96 (5.49) | 47.96 (5.95) | 47.52 (5.98) | 47.10 (4.53) |
| *Post-session* | 50.78 (6.20) | 51.19 (5.10) | 50.33 (7.29) | 50.02 (6.55) | 53.80 (3.26) |
| *Delta* | 3.34 (6.17) | 4.23 (6.02) | 2.38 (6.32) | 2.50 (6.28) | 6.70 (4.57) |
| **Mood: Good/Bad (GB)** | | | | | |
| *Pre-session* | 49.40 (5.70) | 48.96 (5.85) | 49.88 (5.62) | 49.08 (5.96) | 50.70 (4.57) |
| *Post-session* | 51.08 (5.88) | 51.27 (5.34) | 50.88 (6.52) | 49.98 (5.89) | 55.50 (3.27) |
| *Delta* | 1.68 (4.78) | 2.31 (4.84) | 1.00 (4.72) | 0.90 (4.60) | 4.80 (4.39) |
| **Attitude toward VR** | | | | | |
| *Pre-session* | 70.18 (10.73) | 70.65 (10.42) | 69.67 (11.26) | 69.42 (9.79) | 73.20 (14.09) |
| *Post-session* | 75.00 (9.57) | 76.00 (8.90) | 73.92 (10.32) | 73.47 (9.56) | 81.10 (7.16) |
| *Delta* | 4.82 (8.76) | 5.35 (9.35) | 4.25 (8.23) | 4.05 (8.29) | 7.90 (10.33) |

**Table 3.** Pairwise correlation and adjusted 95% CI for all variables.

| | MOCA | Age | Computer Proficiency | Computer Self-efficacy | Nature Exposure | Initial Attitude | Depression |
|---|---|---|---|---|---|---|---|
| **MOCA** | | [-0.48,0.29] | [-0.28,0.49] | [-0.29,0.47] | [-0.42,0.31] | [-0.27,0.29] | [-0.30,0.45] |
| **Age** | -0.11 | | [-0.28,0.35] | [-0.38,0.31] | [-0.51,0.28] | [-0.40,0.31] | [-0.55,0.25] |
| **Computer Proficiency** | 0.12 | 0.04 | | [ 0.22,0.80] | [-0.37,0.30] | [-0.25,0.55] | [-0.28,0.51] |
| **Computer Self-efficacy** | 0.1 | -0.04 | 0.58 ** | | [-0.52,0.27] | [-0.21,0.58] | [-0.44,0.30] |
| **Nature Exposure** | -0.07 | -0.13 | -0.04 | -0.15 | | [-0.53,0.26] | [-0.17,0.61] |
| **Initial Attitude** | 0.01 | -0.05 | 0.18 | 0.22 | -0.16 | | [-0.28,0.50] |
| **Depression** | 0.09 | -0.18 | 0.14 | -0.08 | 0.26 | 0.13 | |
| ***Mean*** | *25.88* | *67.97* | *25.91* | *56.86* | *26.6* | *70.18* | *32.25* |
| ***SD*** | *2.81* | *4.85* | *3.39* | *8.16* | *4.04* | *10.73* | *18.76* |

** = Significant at the 0.001 level after Holm's correction for multiple comparisons.

### *3.2. Changes in Mood States*

The first hypothesis predicted that short-term exposure to the virtual environments would be associated with in improvements in mood states. We calculated the change on the various outcome measures from pre- to post-VR experience per participant and performed t-tests to examine if the change was different from 0. We used Wilcoxon signed rank test for awake/tired since it violated normality assumption (W = .949, p= .030).

The results of the Wilcoxon signed rank test indicated that changes in awake/tired states after the VR environment exposure were not significant (V=613, p=0.283). T-tests found that the good/bad (t(49)=2.486, p=0.016, Cohen's d=.35) and calm/nervous mood dimensions (t(49)=3.826, p<0.001, Cohen's d=.54) were significantly positively enhanced after the VR experience, with medium-small and medium effect sizes respectively (Table 4 and Figure 3).

### *3.3. Changes in Attitude toward VR Technology*

The second study hypothesis predicted that short-term exposure to the virtual environments would be associated with improvements in attitudes toward VR technology. A one-sample t-test confirmed this hypothesis (p <0.001) (Table 4 and Figure 3).

**Table 4.** Wilcoxon and T-tests for changes in mood and attitude.

| | **x̄ (SD)** | **V[1]** | **p** |
|---|---|---|---|
| Mood: AT | | 613 | 0.283 |
| | | **t(49)[2]** | **p** |
| Mood: CN | 3.34 (6.17) | 3.826 | <0.001 |
| Mood: GB | 1.68 (4.78) | 2.486 | 0.016 |
| Attitude | 4.82 (8.76) | 3.892 | <0.001 |

*Note.* AT: awake/tired. CN: calm/nervous. GB: good/bad.

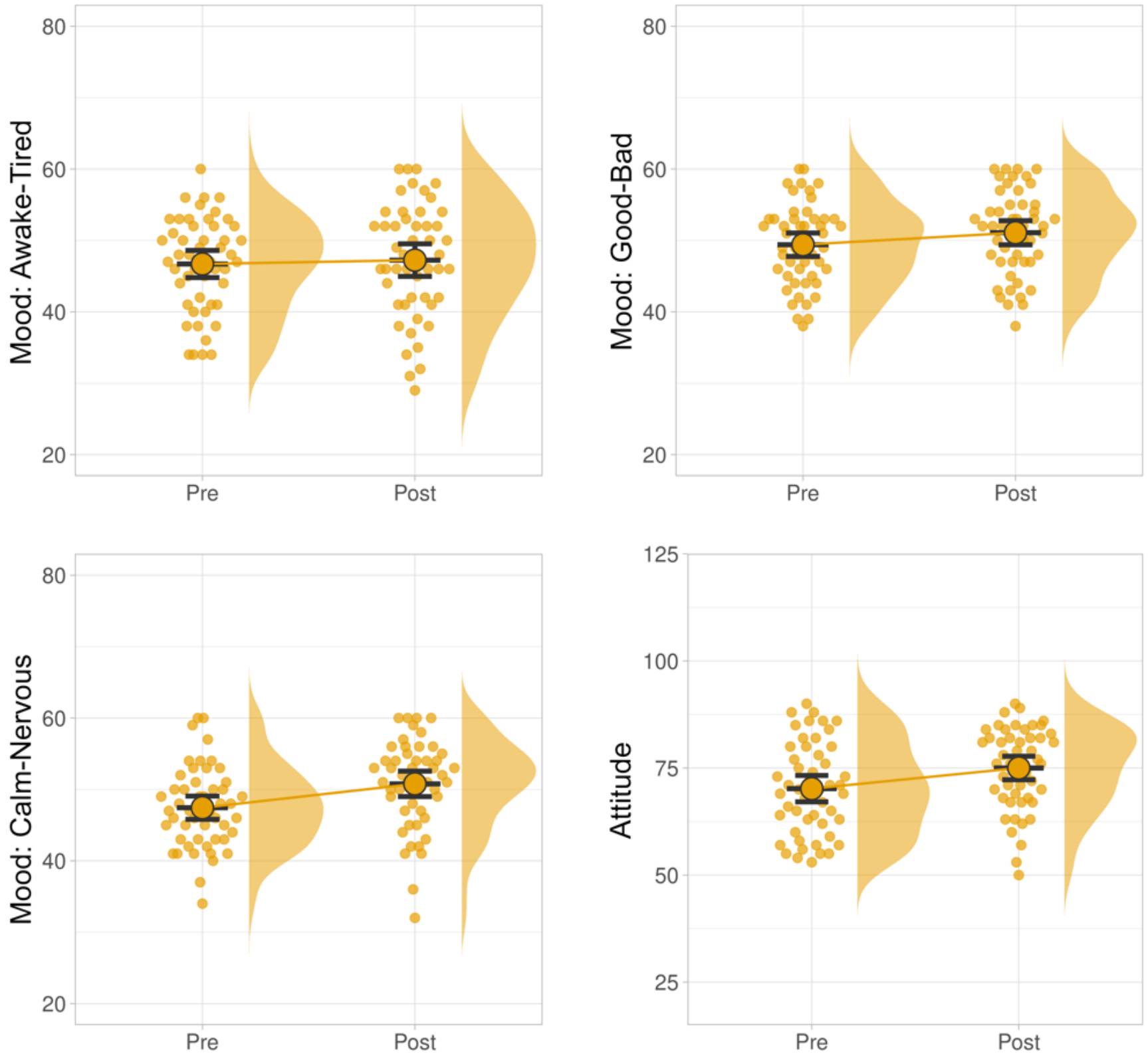

**Figure 3.** Changes in mood states and in attitudes toward VR technology before and after exposure to the virtual garden environment. Significant improvements were seen in all cases except for the Awake–Tired mood state.

### *3.4. CI vs. Non-CI Participants*

The third hypothesis predicted that changes in mood states and attitudes toward VR would not differ significantly between participants with different cognitive abilities. As mentioned in the method section, in this analysis, we used MOCA score to divide the participants into a group with cognitive impairment and a group without cognitive impairment. We fitted a linear model that examined the effects of CI on outcomes while adjusting for Age, Education Level, Computer Efficiency (both CPQ and CSE scores), and Exposure. Then, we estimated the difference in mood states and attitude between CI and non-CI groups based on the model and examined if they are significantly different from 0. We decided to adjust for covariates since CI status is not manipulable. (Table 5 and Figure 4).

We did not find any significant difference in outcomes between CI and non-CI groups.

**Table 5.** Difference between CI and non-CI groups in Outcomes.

| Outcome | Difference | SE | 95% CI | t | p | Cohen's d |
|---|---|---|---|---|---|---|
| **Mood: AT (Delta)** | -2.69 | 2.06 | [-6.86, 1.47] | -1.306 | 0.198 | -0.38 |
| **Mood: CN (Delta)** | -2.19 | 1.85 | [-5.92, 1.54] | -1.184 | 0.243 | -0.344 |
| **Mood: GB (Delta)** | -1.35 | 1.45 | [-4.28, 1.57] | -0.932 | 0.356 | -0.271 |
| **Attitude (Delta)** | -1.41 | 2.45 | [-6.35, 3.53] | -0.575 | 0.568 | -0.167 |
| **Motion Sickness** | -0.12 | 1.06 | [-2.26, 2.02] | -0.114 | 0.91 | -0.033 |
| **Workload** | 0.13 | 0.16 | [-0.19, 0.45] | 0.833 | 0.41 | 0.242 |
| **Desirability** | -0.1 | 0.64 | [-1.39, 1.19] | -0.156 | 0.877 | -0.045 |
| **Spatial Self** | -1.18 | 1.62 | [-4.44, 2.08] | -0.728 | 0.471 | -0.212 |
| **Spatial Action** | -1.04 | 1.36 | [-3.77, 1.70] | -0.763 | 0.45 | -0.222 |
| **UX** | 2.62 | 1.49 | [-0.39, 5.62] | 1.756 | 0.086 | 0.511 |

*Note.* df = 43. Difference: CI - non-CI. Adjusted for age, education level, computer proficiency, computer self-efficacy, and everyday nature exposure.

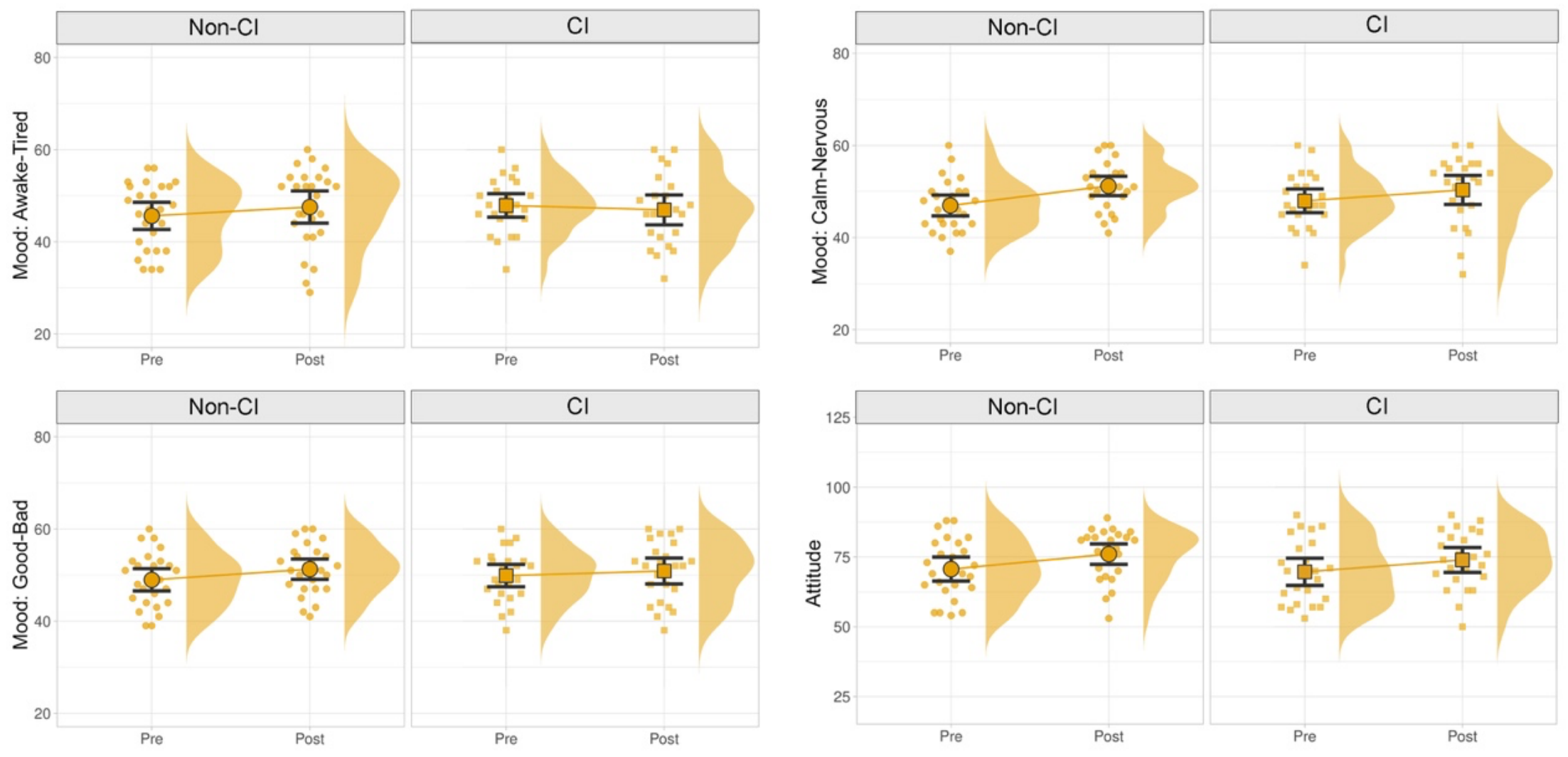

**Figure 4.** Changes in mood states and in attitudes toward VR technology before and after exposure to the virtual garden environment, compared between CI and non-CI groups. There were no significant differences in the changes seen in the CI participants vs. the non-CI participants.

### *3.5.Participants with Disability vs. Those without Disabilities*

The fourth hypothesis predicted that participants with physical disabilities would report more pronounced changes in mood states and attitudes toward VR after exposure to the virtual environment, compared to participants without disabilities. Similar to the examination of CI status in Hypothesis 3, we fitted a linear model that examined the effects of CI on outcomes while adjusting for Age, Education Level, Computer Efficiency (both CPQ and CSE scores), and Exposure. Then, we estimated the difference in mood states and attitude between disability and non- disability groups based on the model and examined if they are significantly different from 0. We decided to adjust for covariates since CI status is not manipulable.

The results indicated a significant difference in the good/bad (GB) mood dimension ($t(43)= 2.244$, $p= .030$) with large effect size ($d=.820$), and a marginal difference in the calm/nervous (CN) mood dimension ($t(43) = 1.821$, $p=.076$) with a medium-large effect size ($d=.665$). In both cases the participants with disabilities showed a larger shift toward positive outlooks, compared to the non-disability group. The awake/tired mood dimension and the attitudes toward VR technology did not show significant differences between participants with disabilities vs. those without disabilities (Table 6 and Figure 5).

**Table 6.** Difference between Disability and non- Disability groups in Outcomes.

| Outcome | Difference | SE | 95% CI | t | p | Cohen's d |
|---|---|---|---|---|---|---|
| **Mood: AT (Delta)** | 2.16 | 2.62 | [-3.13, 7.45] | 0.824 | 0.415 | 0.301 |
| **Mood: CN (Delta)** | 4.15 | 2.28 | [-0.45, 8.74] | 1.821 | 0.076 | 0.665 |
| **Mood: GB (Delta)** | 3.91 | 1.74 | [0.40, 7.43] | 2.244 | 0.030 | 0.820 |
| **Attitude (Delta)** | 3.64 | 3.04 | [-2.49, 9.77] | 1.198 | 0.237 | 0.438 |
| **Motion Sickness** | -0.08 | 1.33 | [-2.76, 2.61] | -0.059 | 0.953 | -0.022 |
| **Workload** | 0.01 | 0.20 | [-0.39, 0.41] | 0.050 | 0.960 | 0.018 |
| **Desirability** | -0.44 | 0.80 | [-2.05, 1.17] | -0.555 | 0.582 | -0.203 |
| **Spatial Self** | 4.47 | 1.93 | [0.58, 8.35] | 2.318 | 0.025 | 0.847 |
| **Spatial Action** | 2.53 | 1.67 | [-0.84, 5.91] | 1.512 | 0.138 | 0.552 |
| **UX** | 1.78 | 1.92 | [-2.09, 5.64] | 0.927 | 0.359 | 0.339 |

*Note.* df = 43. Difference: disability - non-disability. Adjusted for age, education level, computer proficiency, computer self-efficacy, and everyday nature exposure.

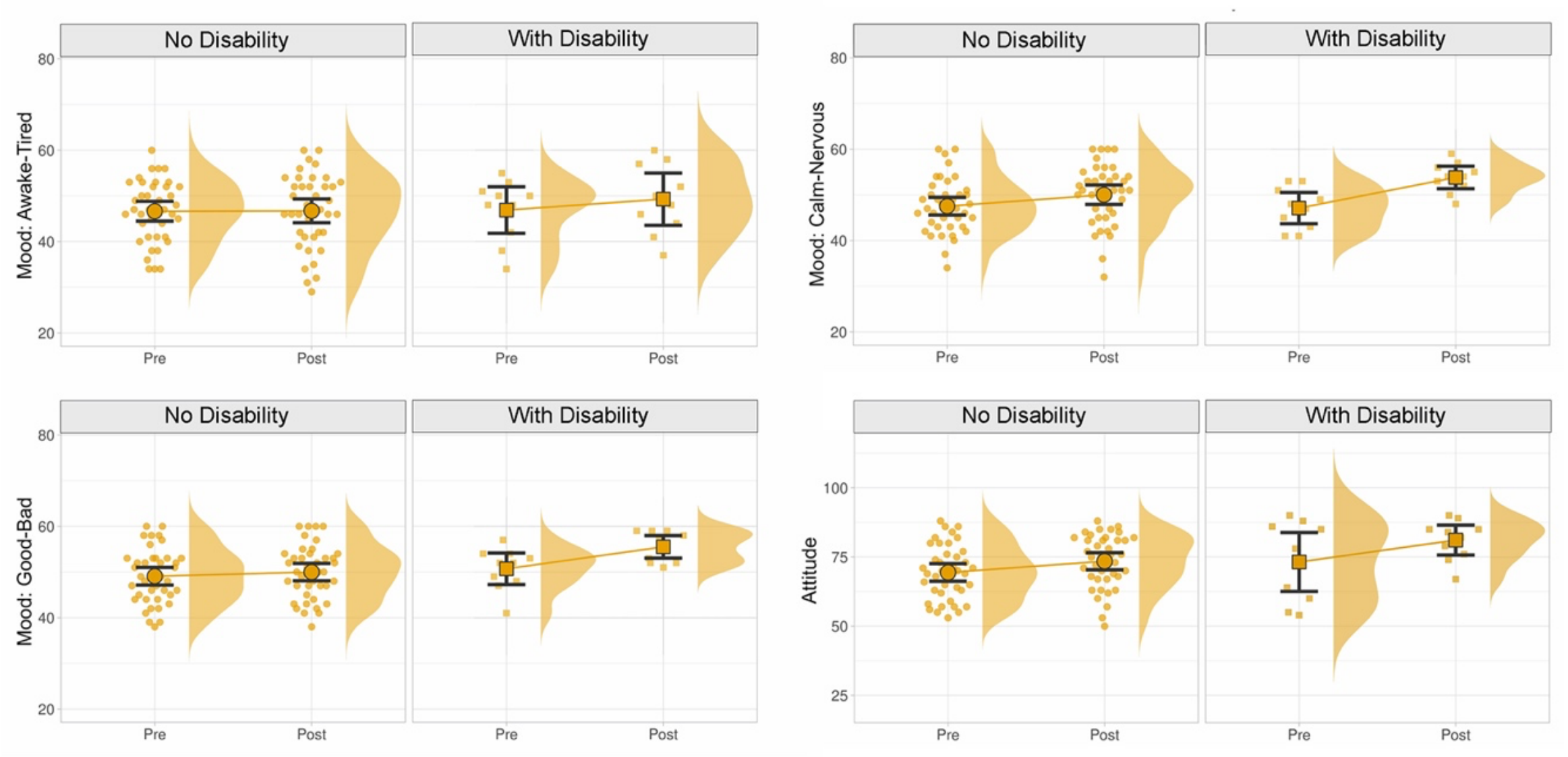

**Figure 5.** Changes in mood states and in attitudes toward VR technology before and after exposure to the virtual garden environment, compared between disability and non-disability groups. Significant differences were found between the two groups for the good/bad and calm/nervous mood dimensions, but not for the awake/tired mood dimension or the attitudes toward VR technology.

### *3.6.Qualitative Results*

All interviews were recorded and transcribed into Word (doc.) files. Content of interview material was analyzed using Lincoln and Guba's (1985) Naturalistic Inquiry strategies for interpreting qualitative data. This method involves the structured analysis of interview transcripts by separating content into discrete segments, sorting the data into categories until saturation is achieved and summarizing the results. The three primary themes that emerged in the interviews were: (a) general qualities of the VR experience, (b) controls and interactive features, and (c) the applicability and impact of the platform for older adults. Overall, positive reactions to the VR environment outweighed the negative comments, with many of the participants expressing strong enthusiasm. For example, "I liked it overall . . . I thought it was really interesting!" (P17). The following discussion includes a somewhat unrepresentative sample of responses, as these concerns and areas of improvement were of particular interest to the researchers.

**Overall VR Experience.** Most of the participants indicated that the equipment and controls for the VR were intuitive and easy to learn, and that the experience overall was relaxing and pleasant. In fact, the most common complaint about the VR experience was that the sessions ended too quickly and did not allow enough time for participants to become fully fluent in using the controls or exploring the environment. Several participants expressed feelings of nervousness about taking actions in the environment or about the perspectives that they encountered, most notably regarding one of the nature videos that included a broad vista: “The one where you felt like you were really high . . . [that] made me really nervous and I was afraid to move about for fear of falling” (P09). Another significant negative reaction that was expressed by more than half of the participants was that the digitally designed garden felt “cartoonish” or artificial after the experience of viewing the nature videos: “It’s that fake, Alice in Wonderland thing, but it’s a fun kind of thing” (P27). In a few cases, participants expanded this negative perception of the digital garden to include the entirety of the virtual experience: “I didn’t at any time feel as though I was actually standing in that particular spot . . . it felt like I was participating in a computer-generated experience” (P51). This reaction was contradicted, however, by the larger number of participants who emphasized the realism of the 360-degree video footage: “I felt like I was really right in it!” (P41). One participant noted that in the past they had frequently visited a site where we filmed one of the nature scenes: “Having stood in that exact spot on the hill it was . . . very, very close to real” (P29).

The participants had mixed reactions in regard to the how immersive the environment felt, with approximately equal numbers expressing a sense of full immersion vs. a sense of being a detached observer or not really present in the environment. The most commonly cited obstacles to immersion were physical reminders of the real-world context, such as when the connecting cables of the headset brushed up against the participant’s body, or when they had to adjust their position due to moving too close to the edge of the VR floor-space. In addition, one participant indicated difficulties in focusing on the display: “[The image] was just slightly out of focus . . . I still felt like I was there and I was just needing to focus my eyes a little differently” (P54). While a few of the participants placed an emphasis on the unreality of the experience, most felt that immersion-breaking moments were exceptions: “[I was] so immersed that I was always surprised when the real world intruded, whether it was the cord or the visual reminder that I was near the wall. . . . It really felt real and wonderful and I’m really crazy about nature. . . . so this this was just amazing. I felt like I was in nature” (P47).

**Controls and Interactive Features.** Participants generally reported overcoming the learning curve quickly enough to enjoy the virtual experience, although they sometimes made mistakes with the hand-held controllers. Of greater concern was that multiple participants commented on issues with holding the controllers, indicating that the joystick was difficult to grasp and/or manipulate. Several individuals volunteered that they thought the controls were too small and would be difficult for some older individuals to use, particular due to the need for precise hand movements: “My hands are pretty arthritic—they don’t hurt and they can still move—but at some point what happens when people’s hands aren’t that mobile?” (P50). Another participant linked the issue of cognitive impairments with potential interface difficulties: “I think that people with mild cognitive impairments might have problems with the controls” (P41). These comments indicate technological design needs that are somewhat beyond the scope of the current study—it seems probable that VR platform manufacturers do not view older individuals as their targeted consumer market. If these technologies are to have therapeutic benefit for older adults, however, then at some point the issue of accessible controls will need to be addressed.

Comments about the selection of video sites, the interactive features of the garden, and the included activities were almost uniformly positive. A majority of the participants expressed appreciation for the sense of closeness to plants and animals and the ability to interact with them in the garden. One participant observed a glitch in the programming that caused the watering can to spontaneously switch between right and left hands, and several others expressed disappointment that they were unable to explore visible areas in the garden that had not yet been fully developed for interaction. Overall, the features of the garden were highly regarded, with participants indicating that they did not feel entirely realistic but that they brought delight and were reminiscent of real-world memories: "I like that within the virtual world I was able to make choices, like picking flowers. I enjoyed the proximity of the flowers and the animals like the duckling" (P14).

**Applicability and Impact of the Platform**. When asked about their willingness to engage in the experience again and their likelihood of recommending it to others, approximately three-quarters of the participants responded positively. Many of the comments in this area focused on the value of the VR environment for individuals who lack the ability to experience real-world nature in person, such as those with severe cognitive impairment or limited mobility: "[The VR experience could] transport them out of the crappy situation they've been dealt, into a beautiful place" (P41). Several participants posited that the VR experience would be especially useful during a pandemic, when opportunities to visit real-world natural areas were limited. A few participants noted that the headset felt heavy and that it might create balance and confusion problems for older adults, particularly those who were taking strong medications. Some of the respondents remained skeptical of VR technology as a general concept, and were conflicted in their responses due to concerns that the VR environment would replace real-life experiences: "It's hard to tell how therapeutic it would be to plant fake flowers in a fake garden. I don't know, I can't quite imagine being in a state where that would be enough, as opposed to just reminding [users] that they couldn't go out into a real garden" (P45). "There's nothing that compares with walking in the real world. . . . [Perhaps if] I had a disability or something and didn't have the opportunity to walk around in the real world. . . . I'm very familiar with the garden, so it was a shallow version of reality" (P32).

The more positive comments about the utility of the VR environment often tended to focus on its role in prompting active engagement, and its potential uses for physical therapeutic purposes: "Absolutely, for people with impairments that are [getting use of] their hands back or whatever, I think it's . . . a wonderful tool" (P43). "I can imagine it . . . keeping you intrigued and keeping you mentally stimulated as well as physically stimulated" (P53). "I think would be wonderful to keep an active mind and I think would be great idea" (P37). Notably, the majority of both positive and negative comments about the applicability of the platform tended to focus on its impacts for *other* users who were imagined to be in different circumstances than the speaker. This may have been due to a hesitation on the part of participants to see themselves as a potential user of virtual environments, even if they had experienced benefits from their encounter with the technology.

## 4. Discussion

Overall, the study findings indicated that the virtual environments were suitable for aging adults with and without a cognitive impairment. The VR experience had an overall positive impact on mood and resulted in improved attitudes toward the technology. Very little cybersickness or frustration with the technology was reported. During the interviews some participants expressed skepticism about the artificialness of the VR. Such individuals were in the minority, however, and their reactions were mostly centered around concerns that the VR might become a replacement for real-world nature.

The general improvements in mood states reported by the study participants after experiencing the VR are consistent with prior findings (Appel et al., 2020; Russell et al., 2013), and they support the theory that VR can serve as a restorative or stress-reducing environment. However, the current study findings do not provide evidence about the long-term impact of VR-use for mental health or cognitive function, which will need to be evaluated in further studies. The study also did not find improvements in the awake/tired dimension of the mood instrument—this may be due to the inherent fatigue involved in learning a new technology and from completing numerous questionnaires during the session. To some extent there may be a trade-off between engagement levels and fatigue, and prior researchers have noted that VR technologies can be intrinsically tiring (Moyle et al., 2018; Siriaraya & Siang Ang, 2014; Thach et al., 2020). These issues need to be accounted for when considering session lengths and frequency of use. However, we did find significant improvements in reporting "good" mood and "calm" mood outcomes after the VR sessions.

The improvements in attitudes toward VR are particularly notable, as older adults frequently exhibit hesitancy and negative feelings about novel technologies (Broady et al., 2010; Hauk et al., 2018). The current study indicated that even one brief session of VR can result in a significant shift in such attitudes toward more positive perceptions. The researchers speculate that further VR sessions would likely result in additional reductions in hesitancy, particularly if the regular use of VR is well-integrated into broader everyday life and the experiences are shared with friends and family (Appel et al., 2020).

The finding that changes in mood and attitude were not significantly different between those with and without a cognitive impairment suggests that these interventions could be beneficial to individuals with declining cognitive capabilities. However, the current study did not evaluate the responses of individuals with more severe degrees of cognitive impairment. There is strong evidence that novel experiences can incite fear and anxiety among older adults with advanced cognitive decline (Moyle et al., 2018), and thus particular caution should be used if introducing such individuals to VR. Adjustments to the experience, such as including human facilitators or guides, may be advisable for those with moderate to severe impairment. In addition, however, we recommend that VR experiences as a therapeutic intervention might be initiated during the early stages of cognitive impairment, which can potentially have the effect of slowing decline while also promoting familiarity and comfort with the environment over time. Reminiscence-based activities have been shown to be effective in improving mood and engagement (Chin, 2007; Hsieh & Wang, 2003; Manera et al., 2016), and indeed, in the interviews for the current study we found that several participants spontaneously remarked on how the environment brought to mind past experiences. Thus, the stability and consistency of the VR platform and a combination of familiar elements or locations with new and stimulating experiences may be important to consider in long-term therapeutic use.

The study findings indicated that virtual environments could be especially beneficial to individuals with mobility restrictions or other physical disabilities. The relatively greater improvements in mood and attitudes toward VR that were reported by participants with disabilities (compared to those without disabilities) may be grounded in the understanding that such individuals often confront severe accessibility issues and have fewer overall opportunities to engage with nature outside of the VR (Stigsdotter et al., 2018). Correspondingly, it also might be conjectured that individuals will gain a relatively greater benefit from VR if their broader circumstances limit their ability to spend time in nature, such as individuals who live in dense urban conditions or in harsh environmental climates. It is notable, however, that the current study did not find a correlation between the extent of everyday nature exposure and improvements in mood after using the VR. This finding may be due to the study participants having a consistently high level of real-world nature exposure, leading to a lack of statistical variability needed to produce a correlation. It may also be the case that the greater benefits reported by disabled participants are more closely related to specific accessibility features and comfort levels in real-world vs. virtual environments, rather than to access to nature in general.

### *4.1. Technological Considerations*

While the majority of the participants indicated that the VR equipment was intuitive, comfortable, and easy to use (reflecting the overall improvements in attitudes toward VR after the experience), there were some specific technological concerns that arose during the interviews. One such concern was related to the visual perspectives presented in the nature videos, which incited feelings of precarity and/or fears of falling for some participants. This is an understandable reaction, as vertigo is a common human experience and injury from falls are a particularly significant concern for older individuals (Fuller, 2000). This issue needs to be taken into account when filming immersive videos for restorative VR use, by ensuring that the perspectives of the videos are well-grounded in a stable and safe vantage point.

In regard to the digitally created interactive garden, many of the participants expressed a sense of disappointment in the level of graphical realism and sharpness, particularly as they had just viewed the more detailed nature videos a few minutes before. One participant noted: "I thought I'd be seeing the real world and be working in that, but I realize how can you do that when you have to create a world where you can pick something up so it can't necessarily be, you know, pictures" (P43). To some extent, the level of realism in digital environments is related to the amount of time and resources that are available to create the renderings. Ongoing improvements in VR technologies, as well as additional time to create more detailed graphical elements for the virtual garden, will likely help to improve these participant reactions and increase the sense of immersion in the environment.

The physical aspects of the head-mounted display and the hand-held controllers were also a topic of concern. Some participants remarked that the headset seemed heavy, and that the cable connections from the headsets to the computer could be a distraction. In addition, one participant indicated that it was uncomfortable to wear glasses along with the headset. Several participants remarked that the small size of the joystick on the hand-held controllers might create difficulties for individuals with arthritis or who had reduced fine motor skills. While the number of participants reporting discomfort with the head-mounted display or hand-held controllers was not high, these considerations should be taken into account when developing VR applications for older adults.

### *4.2. Future Research*

We are currently working to update the virtual platform based on feedback from the study participants, with an emphasis on improving the digital garden with greater realism and a wider array of interactive features. One important short-term goal is to integrate social components into the environment, which we believe will help to enhance engagement and build bridges between the VR and real-world contexts. These features will allow selected friends, family members, caretakers, and peers to work collaboratively and to share and discuss their experiences in the VR. We are also preparing for future longitudinal studies that will evaluate the impact of regular use of the VR environment over time for mental health states such as depression and for the maintenance of cognitive function. This future work will likely separate the 360-degree nature videos and the digital garden into separate trials, in order to better isolate the impacts of these different forms of virtual experience.

## 5. Conclusion

The development of VR technologies has created promising opportunities for individuals to experience immersive natural scenery and engaging, restorative environments, even when their access to such environments in the real world might be limited. The current study evaluated the impacts of brief exposure to a new interactive garden environment and 360-degree nature videos for older adults with a wide range of physical and cognitive capabilities. The findings indicated significant enhancements in mood states and in attitudes toward VR technology after a 30-minute multi-faceted exploration experience. These benefits were similar among participants of different cognitive abilities. They were significantly increased for participants with physical disabilities (compared to those without physical disabilities). The results indicate that VR may provide a cost-effective, non-invasive, and non-pharmaceutical approach for improving the lives of older adults in both clinical and recreational settings. Interview responses provided by the participants after they experienced the VR environments were positive overall, while providing valuable cautions and areas for improvement, such as the need to avoid precarious vantage points in the nature videos and the need for controllers that are more suited for ageing fingers.